\documentclass{llncs}
\usepackage{graphicx}
\usepackage{url}
\usepackage{epstopdf}
\usepackage{bm}
\usepackage{multirow}
\usepackage[export]{adjustbox}
\usepackage{float}
\usepackage{algpseudocode,algorithm,algorithmicx}
\usepackage{longtable}
\usepackage{amsmath,amssymb,amsfonts}
\usepackage{graphicx}
\usepackage{epstopdf} 
 
\begin{document}

\title{Automated Remote Patient Monitoring: Data Sharing and Privacy Using Blockchain}

\author{Gautam Srivastava\inst{1,4}, Ashutosh Dhar Dwivedi\inst{1,2}, Rajani Singh\inst{2,3}}
\institute{Department of Mathematics and Computer Science, Brandon University, Brandon, Manitoba, Canada \and Institute of Computer Science, Polish Academy of Sciences, Warsaw, Poland \and Faculty of Mathematics, Informatics and Mechanics, University of Warsaw, Warsaw, Poland \and Research Center for Interneural Computing, China Medical University, Taichung, Taiwan, Republic of China}
\maketitle


\section*{Abstract}
The revolution of Internet of Things (IoT) devices and wearable technology has opened up great possibilities in remote patient monitoring. To streamline the diagnosis and treatment process, healthcare professionals are now adopting the wearable technology. However, these technologies also pose grave privacy risks and security concerns about the transfer and the logging of data transactions. One solution to protect privacy in healthcare is the use of blockchain technology. However, one of the primary problems with blockchain is its highly limited scalability. In this work here, we propose the utilization of a  blockchain based protocol to provide secure management and analysis of data. In this paper we use recently introduced PoW based protocol GHOSTDAG, that generalizes Satoshi's blockchain in \cite{Bitcoin} to a direct acyclic graph of blocks (blockDAG) and provides high throughput while also avoiding the security-scalability problem. We use two blockchains based on the original GHOSTDAG protocol, one that is private and one that is public. Using a private blockchain, we create a system where we use smart contracts to analyze patient health data. If the smart contract for any reason issues an alert for an abnormal reading then the system makes the record of that event to the public blockchain. This would resolve the privacy and security vulnerabilities associated with remote patient monitoring and also the limited scalability problem of Satoshi's original blockchain.  

\section{Introduction}
\begin{figure}[ht]
    \centering
    \scalebox{.70}{\includegraphics{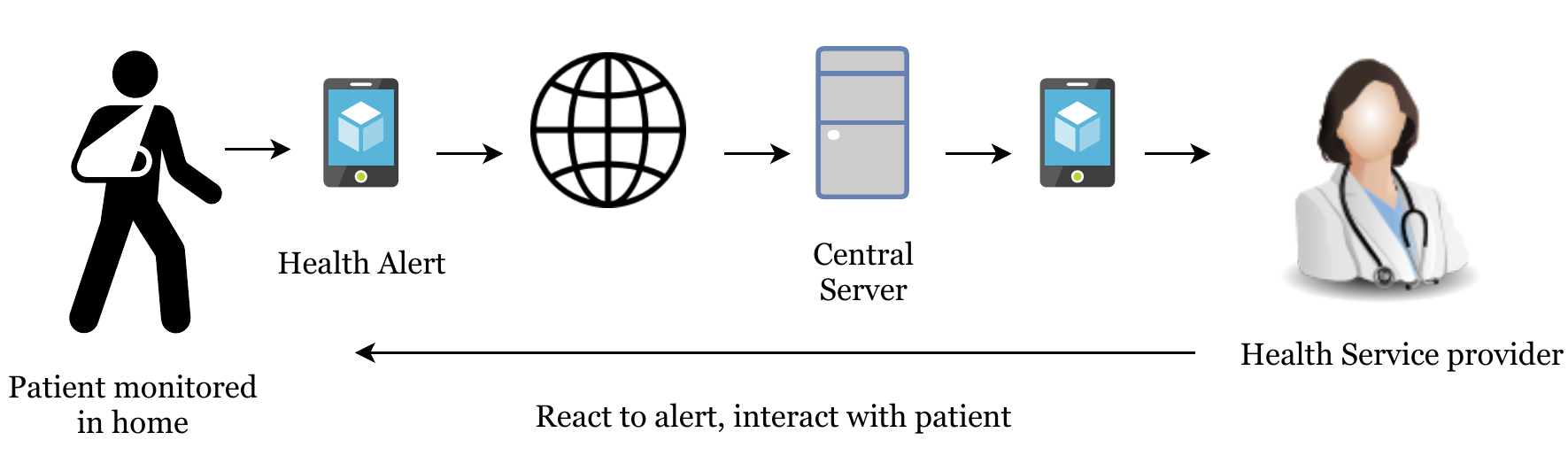}}
    \caption{RPM} 
      \label{RPM}
  \end{figure}

Many countries are suffering from a dramatic increase in the number of patients, and it is becoming more difficult for patients to access primary doctors or caregivers. In recent years, the rise of IoT and wearable devices has improved the patient quality of care by remote patient monitoring. It also allows physicians to treat more patients. Remote patient monitoring (RPM) provides monitoring and care of patients outside conventional clinical setting (in the home as an example). First off, it allows patients an intrinsic convenience of service. Patients can stay connected with health providers as required. It also reduces medical costs and improves the quality of care. This is the main reason that healthcare providers are exploring means by which to provide remote patient monitoring to the masses. The main component of a remote patient monitoring system could be, a specially designed monitoring device to monitor and transmit health data to smart contracts, a smartphone with internet connectivity and a remote patient monitoring app (Fig:\ref{RPM}). Wearable devices and IoT plays an important role in remote patient monitoring. Wearable devices collect patient health data and transfer it to hospitals or medical institutions to facilitate health monitoring, disease diagnosis, and treatment.

\begin{figure}[h!]
	\centering
    \scalebox{.60}{\includegraphics{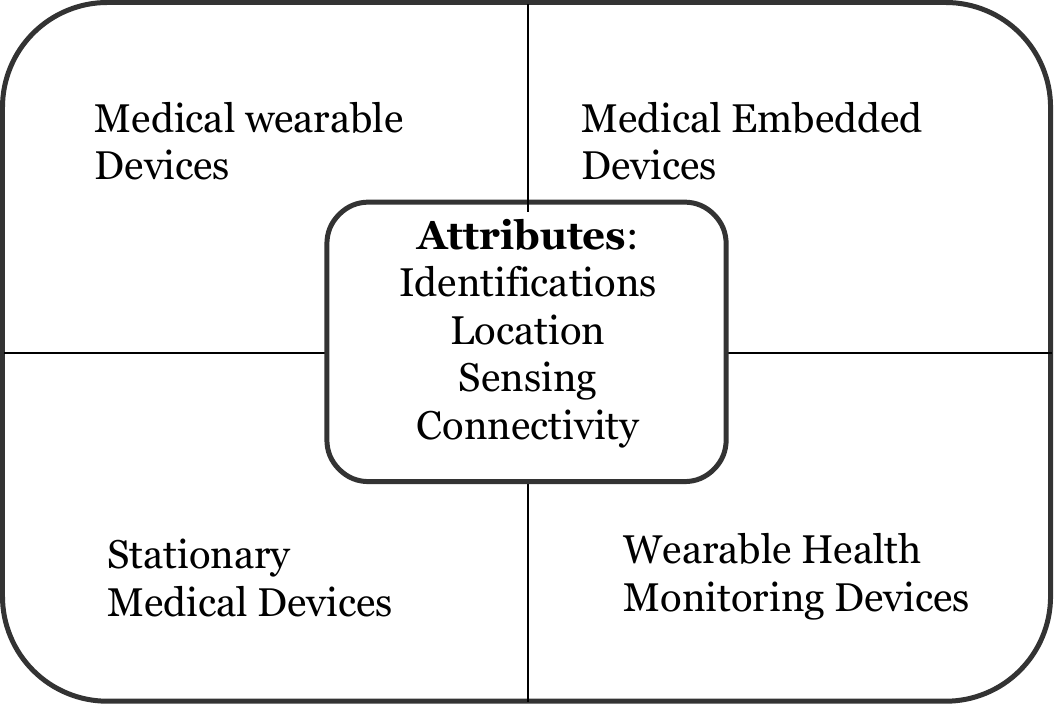}}
    \caption{Healthcare IOT Typology} 
      \label{Typology}
  \end{figure}

Wearable devices in healthcare are the smart electronic devices with micro-controllers that can be embedded into clothing or worn on the body as accessories. They are unobtrusive, user-friendly and connected with advanced features such as wireless data transmission, real-time feedback and alerting mechanisms built into the device. These devices can provide important information to healthcare providers such as blood pressure, blood glucose levels and breathing patterns just to name a few. Healthcare devices can be categorized into four types (Fig:\ref{Typology}): 

\begin{itemize}
\item Stationary Medical Devices - devices can be used on a specific physical location (e.g., chemotherapy dispensing stations for home-based healthcare)
\item Medical Embedded Devices - devices which can be implanted inside the body (e.g., pacemakers)
\item Medical Wearable Devices - prescribed devices by doctors (e.g., insulin pump)
\item Wearable Health Monitoring Devices -  consumer products (e.g., Fitbit, Fuelband, etc.)
\end{itemize}

On November 13, 2017, the Food and Drug Administration (FDA) approved the first pill with a sensor inside of it (aripiprazole tablets with sensor) that can track if a patient has swallowed it. This pill's sensor sends messages to a wearable patch, and the patch itself transmits the message to a mobile application on the smartphone. This technology could be a gamechanger for chronic disease and mental health disorders. 

\begin{figure}[h!]
    \centering
    \scalebox{.70}{\includegraphics{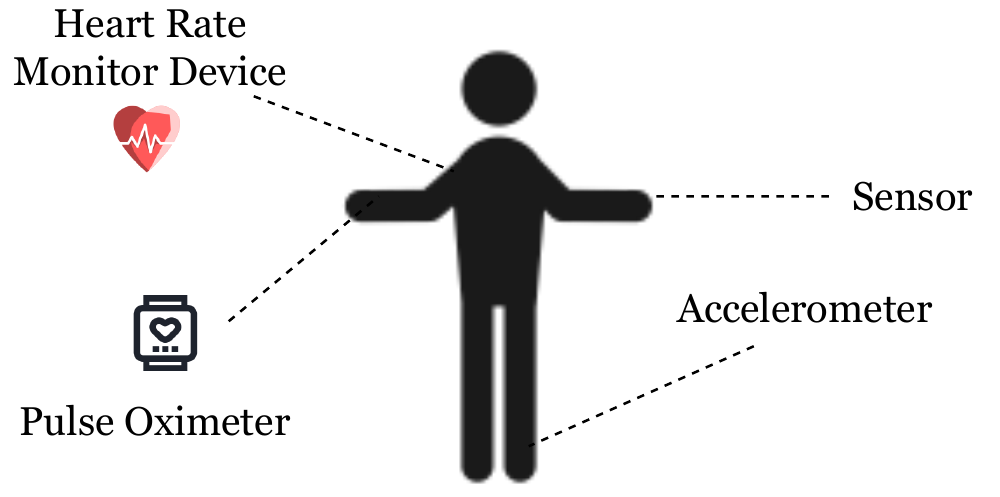}}
    \caption{Wearable devices for the patient } 
      \label{Wearable_Device}
  \end{figure}

One of the facets of the \emph{Internet of Things} (IoT) is the network of wearable devices, embedded with software, electronics, sensors, actuators, and connectivity which enables the wearable device to connect and exchange data (Fig:\ref{Wearable_Device}). To handle such patient data with other institutions, such infrastructure demands secure data sharing. Health data is highly , and sharing of data may raise the risk of exposure. Second, the current system of data sharing uses a centralized architecture which requires centralized trust. 

The solution for data privacy and security could very well be Blockchain technology, initially proposed by Satoshi Nakamoto in \cite{Bitcoin}, blockchain technology provides the robustness against failure and data exposure. It acts as a decentralized architecture to record the data. The inability to delete or change information from blocks makes the blockchain the best appropriate technology for healthcare system. However, questions surrounding scalability and security of blockchain still need to be answered. The security and scalability of blockchain rely on blocks propagating quickly to all miners. When a miner (responsible node for maintaining the blocks) extends a new block to the blockchain, it propagates to all honest nodes before the next one is created. The propagation of such long blockchains brings the problem of scalability and slow computational speed.    
Therefore, here we introduce a more advanced and scalable blockchain system for the remote patient monitoring system. We are using the GHOSTDAG protocol \cite{phantom} --- a protocol for transaction confirmation that is secure under any throughput that the network can support. Instead of long blockchains like its predecessor, GHOSTDAG incorporates all blocks in the form of a Directed Acyclic Graph, a \textbf{blockDAG}. 

The Health Insurance Portability and Accountability Act (HIPAA), which was enacted on $21$ August 1996, set the standard to protect patient health data. The goal of this act is to protect the privacy of patients health information while data is transferred to ``covered entities", i.e., healthcare clearinghouses, health plans or healthcare providers. HIPAA Security Rules outline national security standards to protect health data transmitted electronically. No confidential information or health data of the patient is stored in the blockchain or smart contracts because of HIPAA compliance reason. We only record the fact that event occurred and used the blockchain technology as the ledger.

\section{Related Work} 
In this paper, we pull some of our main motivation to explore blockchain in healthcare from \cite{mettler2016}, who systematically mentioned some of the latest trends in this area of research. Since the introduction of Bitcoin in \cite{Bitcoin}, the possibilities are endless of how the underlying technology can be used in other ways outside of the financial realm.

In \cite{prisco2016}, Gem, a provider of enterprise blockchain solutions, launched Gem Health, a network for developing applications and shared infrastructure for healthcare powered by the Ethereum blockchain, and announced that Philips Blockchain Lab, a research and development center of healthcare giant Philips, is the first major healthcare operator to join the Gem Health network.

In \cite{irving2016}, there was an introduction to methods for using blockchain to provide proof of pre-specified endpoints in clinical trials. Irving and Holden empirically tested such an approach using a clinical trial protocol where outcome switching had previously been reported. They confirmed the use of blockchain as a low cost, independently verifiable method to audit and confirmed the reliability of scientific studies.

Many different healthcare applications were introduced in \cite{angraal2017}, most notably the work in Estonia. Previously Estonia was shown to have used blockchain technology in voting systems \cite{Estonia}. In \cite{angraal2017}, there was reference made to Estonia being able to secure over one million health records using blockchain technology.

In \cite{nichol2016}, Nichol highlights where blockchain is emerging with the potential to transform the patient experience from payers to providers to patients embracing blockchain to create sustainable competitive advantages. Nichol magnifies the principles and uses cases pioneering a new frontier to revolutionize the healthcare experience. Based on his articles, blogs, and musings, the book shows what is required to transform healthcare from the inside.

Finally, in \cite{taylor2016}, Taylor introduces a new project that aims to use blockchain technology to improve the security of the pharmaceutical supply chain. The project which is still at the planning stages envisages using blockchain tracking and time stamps to make it easy to establish exactly when and where medicine was produced.

\section{Drawbacks and security issues} 
The main concern in remote patient monitoring systems is the secure and efficient transmission of the medical data. Healthcare data is a lucrative target for hackers and therefore securing protected health information (PHI) is the primary motivation of healthcare providers. Healthcare has become the primary target for cybercriminals. For example, cyber attacks on medical devices or health data have become more common in the last decade. However, the inability to delete or change information from blocks makes blockchain technology the best technology for the healthcare system and could prevent these issues. However, using blockchain technology in its original form is not enough of a solution because of its slow computational speed.

\section{Our System} 
We break down our system into five entities:

\begin{itemize}
\item patient
\item healthcare provider
\item wearable devices
\item GHOSTDAG protocol network
\item insurance company.
\end{itemize}

We describe our system in detail as follows:

\begin{enumerate}
\item{Wearable and smart devices:}
Wearable health devices are used by patients to provide health information. These devices are connected by smartphones and health data in raw form and are transferred to smartphones. These innovative devices will be able to measure glucose levels without ever drawing a drop of blood, detect breast cancer through an implant worn in cloths or can detect heartbeats.

\item{Patient:}
The system will collect all health data from the patient. Such data could be heartbeats, sleeping conditions, or walking distance to name a few. Patients themselves are the owners of their personal data and responsible for granting, denying or revoking data access from any other parties, such as insurance company or health care providers. If the patient needs medical treatment, he or she will share personal health data with the desired doctor. Once the treatment is finished the patient can deny further access to the doctor, healthcare provider or health insurance company.  
  
\item{Healthcare provider:}
Healthcare providers are appointed by insurance companies or by patients to perform medical tests or to provide medical treatments. Healthcare providers can request directly to the patient to access previous health data and medical treatment history. Healthcare services provide treatment to patients once they receive an alert from smartphones.

\item{GHOSTDAG protocol network:}
We use two blockchain based protocols, private blockchain where we perform all experiment with patient health data using smart contracts and if smart contracts issue an alert we write the event to public blockchain and send an alert to smart devices and hospitals. These blockchains are based on GHOSTDAG protocol described in section (\ref{GHOSTDAG}).

\item{Insurance company:}
A patient may request to his/her health insurance company to provide healthcare services or to provide good health insurance quote for future. To provide the best medical facilities or healthcare, the insurance company may request data access from users including user health data from wearable devices and medical treatment history. The insurance claim events can also be stored in the blockchain. However, we have not added this entity in out block diagram as insurance is optional depending on patient choice.
\end{enumerate}

\subsection{Proposed System}
We propose a model that does not replace the present blockchain remote patient monitoring system but integrates a new technology right into the current system. 

\subsection{System Requirements} 
\begin{enumerate}
\item{Authentication:} 
Only authorized entities can access the blockchain for health or treatment data. In order to maintain an accurate timeline of events and protect the integrity of the patient’s care, IoT devices like smart contracts must be securely logged-in. Only health care providers and insurance companies permitted directly by the patient can access the health or treatment data.

\item{Scalability and Security:} Medical devices or health data must be accurate and cannot be changed by hackers. The infeasibility to delete or change information from blocks in the chain makes blockchain the best technology for healthcare systems. However, blockchain alone is not enough of a solution because of its slow computational speed. For this purpose, we are using the GHOSTDAG protocol which is more secure and significantly faster than the original blockchain protocol. 
\end{enumerate}

\subsection{GHOSTDAG protocol} \label{GHOSTDAG}
The security of the blockchain protocol depends on the speed of block propagation to all miners in the network before the new block is created. Each new block creation also requires a proof-of-work and therefore block creation itself is a slow process. For a blockchain to be secure, block propagation time must be faster than the time taken by the network together to create a new block. Blockchain suffers from a highly restrictive throughput on the order of $3$-$7$ transactions per second.

\begin{figure}[ht!]
    \centering
    \scalebox{.70}{\includegraphics{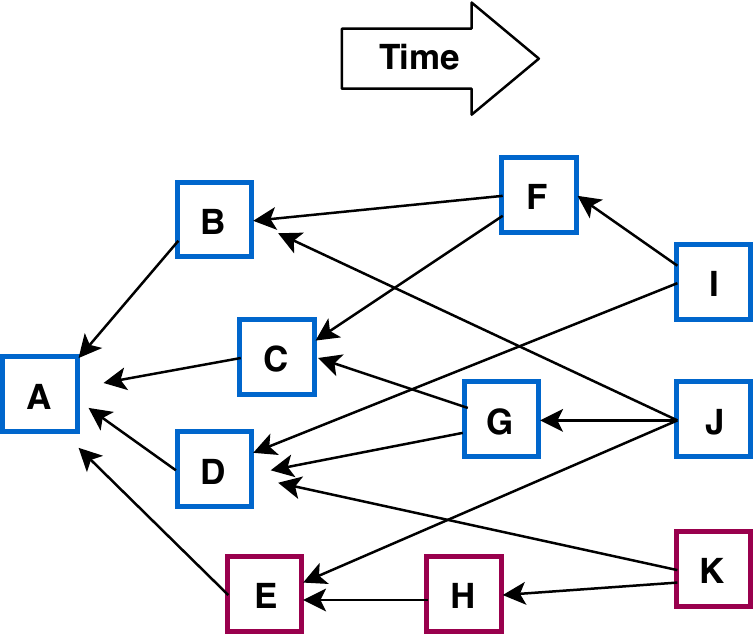}}
    \caption{3-cluster of block DAG } 
      \label{3-cluster}
  \end{figure}

In contrast, the GHOSTDAG protocol, a variant of the PHANTOM\cite{phantom} protocol (but more suitable for practical implementation of blockchain) is a generalization of Satoshi's longest chain protocol. Instead of a long chain of blocks, the GHOSTDAG protocol structures blocks in the form of a Directed Acyclic Graph, a blockDAG. GHOSTDAG structures the blocks in a $k-$cluster (Fig:\ref{3-cluster}), which includes coloring of blocks as Reds (blocks outside the cluster) and Blues (blocks inside the cluster). 

An example shown in Figure \ref{3-cluster} represents the largest $3$-cluster of blocks within a given DAG: $A, B, C, D, F, G, I, J $ (coloured blue). Each block in the cluster has at most $3$ blue blocks in the anticone. On the other hand for blocks $E, H, K$ (coloured red) have more than $3$ elements in anticone.  

By setting the parameter $k=3$, at most $4$ blocks can be created in a single unit of time. The only difference between PHANTOM and GHOSTDAG is, instead of searching for longest $k$-cluster, GHOSTDAG finds a cluster using a greedy algorithm. The maximum $k$-cluster problem is NP hard, and this is the reason PHANTOM is less practical for an ever-growing blockDAG while GHOSTDAG uses a greedy algorithm that is more suitable for implementation. 

\begin{figure}[h!]
    \centering
    \scalebox{.70}{\includegraphics{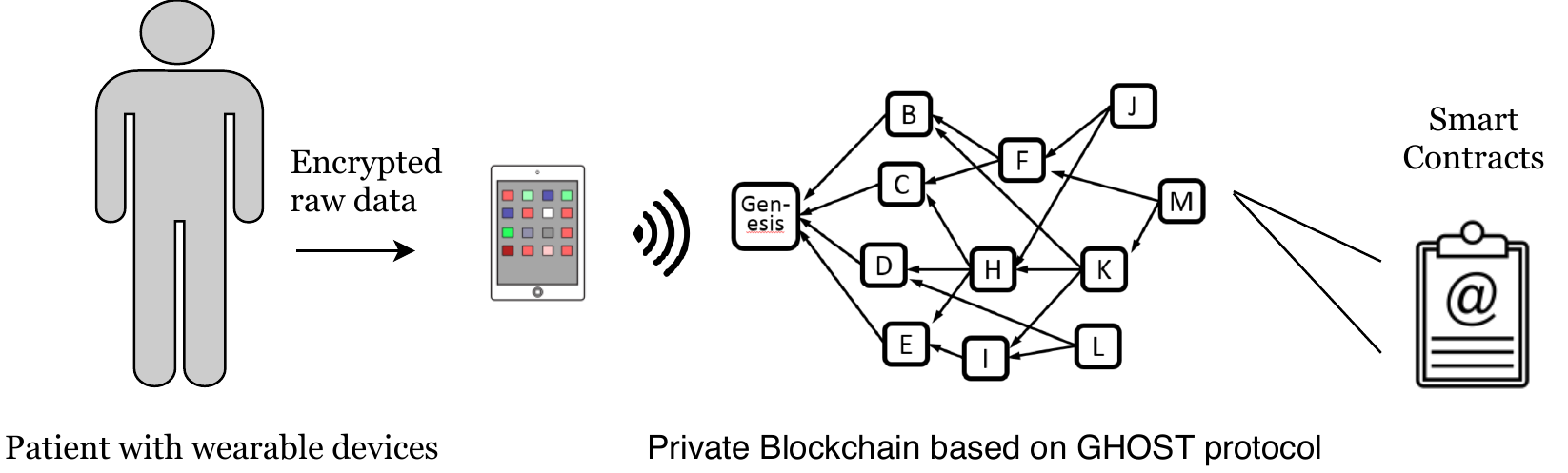}}
    \caption{Raw health data is aggregated by the mobile device and then sent to nodes in the blockchain for processing by the smart contract } 
      \label{PrivBC}
  \end{figure}

\section{Implementation}
In our system, the patient is equipped with wearable devices such as a blood pressure monitor, insulin pump, or other known devices. The health information is sent to the smart devices such as a smartphone or tablet for the formatting and aggregation by the application. Once complete, the formatted information is sent to the private blockchain to the relevant smart contract for full analysis along with the threshold value (Fig. \ref{PrivBC}). The threshold value decides whether the health reading is NORMAL as per standard readings or not. 

  \begin{figure}[h!]
	\centering
    \scalebox{.70}{\includegraphics{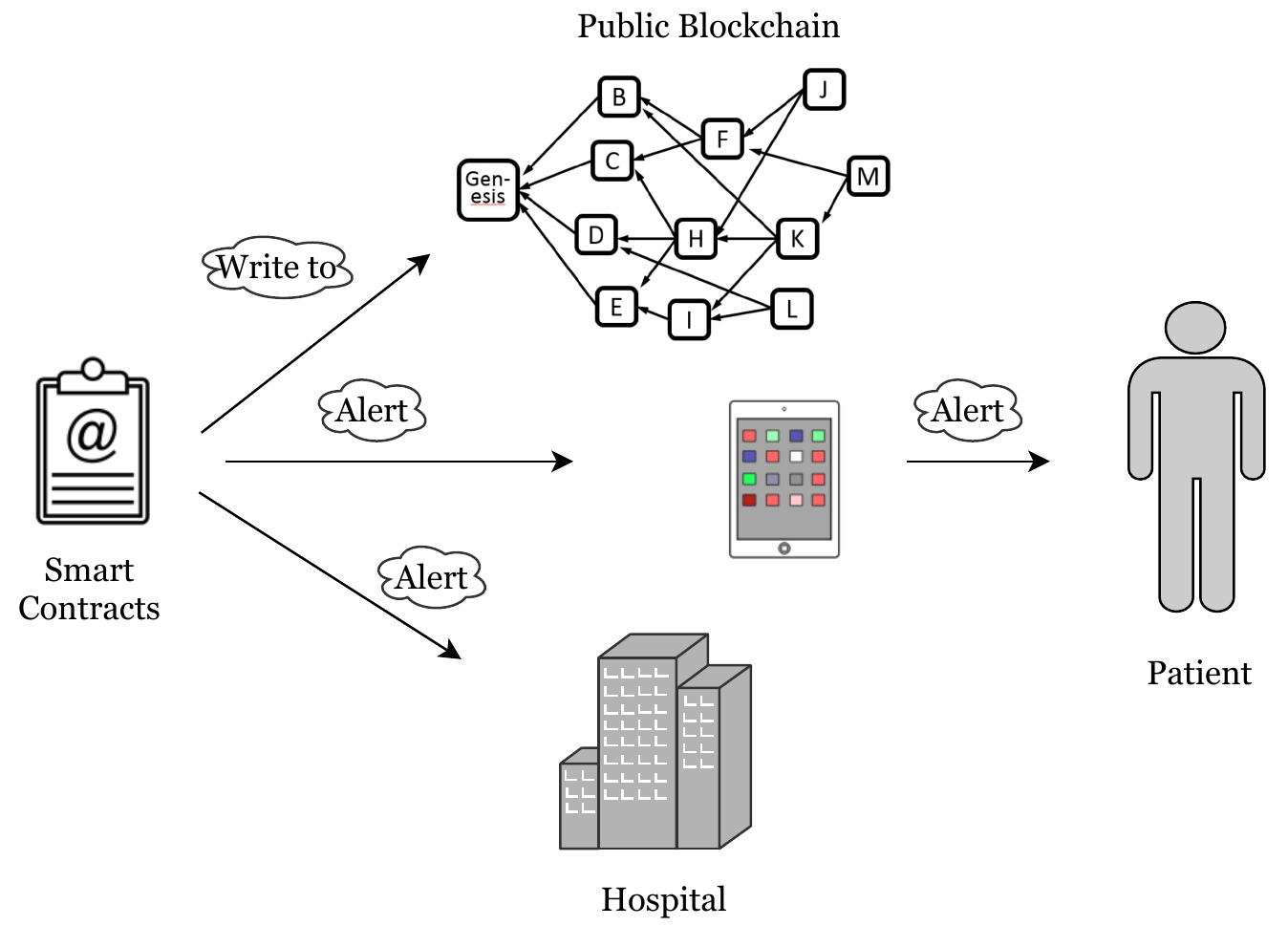}}
    \caption{Smart contracts issue an alert only for an abnormal readings and write records of that event to the public blockchain } 
      \label{PubBC}
  \end{figure}
  
If the health reading is abnormal, then the smart contract will create an event on public blockchain and send an alert to smart devices and hospitals  (Fig. \ref{PubBC}). We can use smart contracts such as Oracle \cite{ConsenSys} that can directly communicate to Oracle smart devices or smartphones.

   \begin{figure}[h!]
    \centering
    \scalebox{.70}{\includegraphics{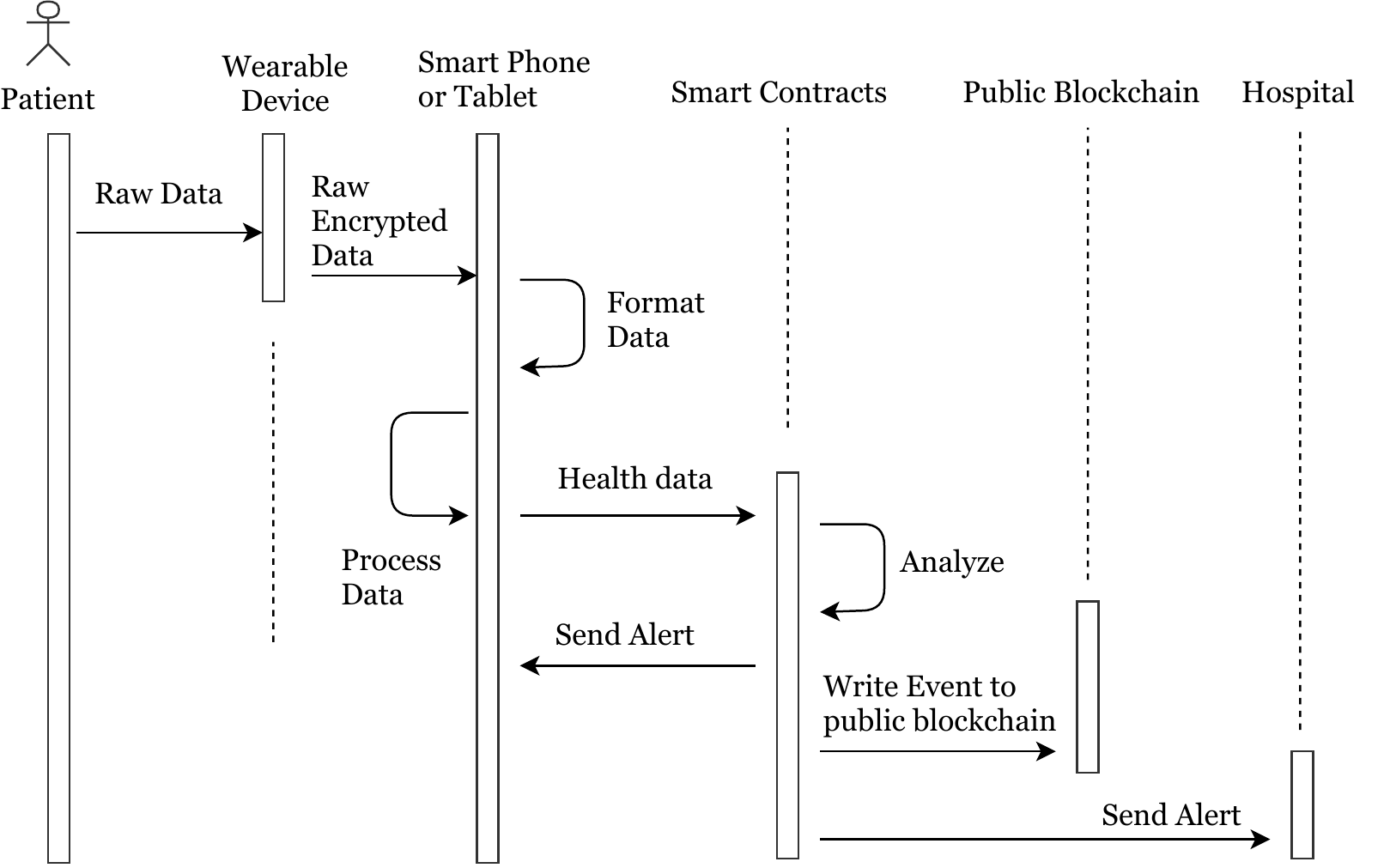}}
    \caption{Logical flow execution of the system } 
      \label{flow}
  \end{figure}

However, we do not store confidential medical information in smart contracts or on the blockchain because of HIPAA compliance reason. Public blockchain will only store the event when an alert is issued. The measurement of health data by wearable devices will be forwarded to a designated Electronic Health Record (EHR) storage. Also, the treatment commands from smart contracts or hospitals will be forwarded to EHR storage while the transaction event is stored in the blockchain. The blockchain transactions are linked with EHR to provide authentication of data in patient medical history. It helps to prevent and detect alteration of patient data in EHR. Designated nodes can only execute, and smart contracts.  These designated nodes are also responsible for verifying the new block. Limiting the viewers such as care providers, device manufacturers, and patients themselves will reduce data exposure of information.  
  
\section{Conclusion} 
We proposed utilizing GHOSTDAG blockchain-based smart contracts to perform real-time health data analysis of patients. The system uses smart contracts that trigger alerts for the healthcare and patient when appropriate and also records the details of transactions on the blocks using the GHOSTDAG protocol. Our model automates the delivery of health-related notifications in a HIPAA compliant manner. We do not store patient health information in the blockchain as putting entire health records onto a blockchain would greatly inflate the size of the entire chain, which would then require much more storage at each node. Therefore, the health data is transferred to EHR (Electronic health record). We are only recording the fact that events occurred using the blockchain technology as a ledger and linked transactions with EHR to provide authentication of data in patient medical history. This, in turn, will help to prevent and detect alteration of patient data in EHR. When slow computational speed and energy consumption are major problems, our model provides secure, high throughput and fast remote patient monitoring system compared to traditional blockchain remote patient monitoring. 

\section{Compliance of Ethical Standards}

\subsection{Funding}
The work of Ashutosh Dhar Dwivedi and Rajani Singh is funded by Polish National Science Centre, project DEC-2014/15/B/ST6/05130 and\\ project 2016/21/N/HS4/00258, respectively. 

\subsection{Conflict of Interest}
Dr. Gautam Srivastava declares that he has no conflict of interest. Ashutosh Dhar Dwivedi declares that he/she has no conflict of interest. Rajani Singh declares that she has no conflict of interest.

\subsection{Ethical Approval}
This article does not contain any studies with human participants or animals performed by any of the authors.

\bibliographystyle{splncs03}
\bibliography{biblio}

\end{document}